\documentstyle[preprint,prb,aps,fleqn]{revtex}
\begin{document}
\draft
\title{Phase transitions in Ising magnetic films and superlattices}
\author{Xiao-Guang Wang\thanks{e-mail:xyw@aphy.iphy.ac.cn}}
\address{Laboratory of optical physics, Institute of Physics,
 Chinese Academy of Sciences, Beijing 100080, People's Republic of China}
\author{Shao-Hua Pan and Guo-Zhen Yang}
\address{China Center of Advanced Science and Technology (World Laboratory),
 P.O.Box 8730, Beijing 100080,People's
 Republic of China and Institute of Physics,Chinese Academy of Sciences,
 Beijing 100080, People's Republic of China}
\author{Received at June 8, 1999 and accepted at July 8,1999 by Z.Z.Gan}
\maketitle

\begin{abstract}
Within the framework of mean field theory, we examine the phase transitions in Ising
magnetic films and superlattices. By transfer matrix method, we derive two general nonlinear equations for phase transition temperatures of Ising magnetic films and superlattices,
respectively. The equations can be applied to the films and superlattices
with arbitrary exchange interaction
constants and arbitrary layer number. Numerical results for
phase transition temperatures as a function of exchange interaction constants are presented.
\end{abstract}
\pacs{Keywords:A.Phase transitions, B.Ising model.}

\section{Introduction}
Considerable effort has been recently devoted to the
understanding of magnetic films,
layered structures and superlattices[1-7]. With the development of molecular beam epitaxy, it is now
possible to grow in a very
controlled way magnetic films with few atomic layers or even monolayer atop nonmagnetic substrates. A superlattice in which the atoms vary from one monolayer
to another can also be envisaged. Very often one
finds unexpected and interesting properties in these systems.
For example, experimental studies[8-10] on the magnetic
properties of surfaces of Gd,Cr and Tb have shown that a
magnetically ordered surface can coexist with a magnetically
disordered bulk phase.

Most work has been devoted to the free surface problem in which
spins on the surface interact with each other through an exchange
interaction $J_s$ different from that of bulk material[11-20].
For the values of $J_s/J$ above a certain critical
value $(J_s/J)_{crit}$ the system orders on the surface before
it orders in the bulk. Below this critical value only two phases
are expected, namely the bulk ferromagnetic and paramagnetic phases.

Magnetic excitations in superlattices were studied in
numerous papers (see, e.g., Ref.21 for a brief review).
Yet less attention has been paid to critical behavoir.
The phase transition temperatures for a Heisenberg magnetic
superlattices have been studied [22-24]. Hinchey and Mills
have investigated a superlattice structure with alternating
ferromagnetic and antiferromagnetic layers[25-26].

In this paper, we are concerned with phase transitions
in Ising magnetic films and superlattices.
we study them within the framework of mean field theory.
The transfer matrix mathod is used to derive nonlinear
equations
for magnetic Ising films and superlattices.
The equations are general ones for arbitrary
exchange interaction constants. In section II,
we outline the formalism and derive the nonlinear
equation for phase transition temperatures of magnetic films.
The transition
temperatures as a function of surface exchange
constants are studied. The nonlinear equation for
transition temperatures of superlattices are given
in section III. The last section IV is devoted to a
brief discussion.

\section{Formalism and phase transitions in Ising magnetic films}
We start with a lattice of localized spins with spin equal to $1/2$. The
interaction is of the nearest-neighbor ferromagnetic Ising type. The Ising
Hamiltonian of the system is given by
\begin{equation}
H=-\frac{1}{2}\sum_{(i,j)}\sum_{(r,r^{\prime})}J_{ij}
S_{ir}S_{jr^{\prime}},
\end{equation}
where
$(i,j)$ are plane indices, $(r,r^{\prime})$ are different
sites of the planes, and $S_{ir}$ is spin variable.
$J_{ij}$ denote the exchange constants and are plane dependent.
We will keep only nearest-neighbour terms.

In the mean-field theory, $S_{ir}$ is replaced by its mean value $m_i$
associated with each plane, and is determined by a set of simultaneous equations
\begin{equation}
m_i=\tanh[(z_0J_{ii}m_i+zJ_{i,i+1}m_{i+1}+zJ_{i,i-1}m_{i-1})/K_B T],
\end{equation}
where $z_0,z$ are the numbers of nearest neighbours in the place and between the
planes, respectively, $k_B$ is the Boltzman constant and $T$ is the
temperature.

Near the transition temperature, the order parameter $m_i$ are small,
Eq.(2) reduces to
\begin{equation}
K_BTm_i=z_0J_{ii}m_i+zJ_{i,i+1}m_{i+1}+zJ_{i,i-1}m_{i-1}.
\end{equation}

Let us rewrite the
above equation in matrix form in analogy with Ref.[27] 
\begin{equation}
{m_{i+1} \choose m_i}=M_{i-1} {m_i \choose m_{i-1}}
\end{equation}
with $M_{i-1}$ as the transfer matrix defined by 
\begin{equation}
M_{i-1}=\left(\matrix {(K_BT-z_0J_{ii})/(zJ_{i,i+1})&
-J_{i,i-1}/J_{i,i+1}\cr 1&0} \right).
\end{equation}

We consider a magnetic film which contains
$N$ layers with layer indices $i=1,2...N$.
From Eq.(4), we get
\begin{equation}
{{m_N}\choose {m_{N-1}}}=R{m_2\choose m_1},
\end{equation}
where $R=M_{N-2}...M_2M_1$ represents
successive multiplication of the transfer
matrices $M_i$.

For an ideal film system,
there exists symmetry in the direction perpendicular
to the surface, which allows us to write $m_i=m_{N+1-i}$.
Then, the following nonlinear equation for
the transition temperature can be obtained from equations
(3) and (6) as
\begin{equation}
R_{11}[(K_BT-z_0J_{11})/(zJ_{1,2})]^2+(R_{12}-R_{21})
[(K_BT-z_0J_{11})/(zJ_{1,2})]-R_{22}=0.
\end{equation}

The above equation is the general equation for
the transition temperature of symmetric films.
It is suitable for arbitrary exchange interaction
constants $J_{ij}$. All the information about the phase transition temperatures of the 
system is contained in the equation.

For a uniform system with $J_{ij}=J$, Eq.(7) reduces to
\begin{equation}
R_{11}(t-z_0/z)^2+(R_{12}-R_{21})(t-z_0/z)-R_{22}=0,
\end{equation}
where $t=K_BT/(Jz)$ is the reduced temperature and $R=D^{N-2}$. Here
the matrix
\begin{equation}
D=\left(\matrix{t-z_0/z & -1 \cr 1 &  0}\right).
\end{equation}

Note that $\det(D)=1$, we can linearalize  matrix $R$ as[28]
\begin{equation}
R=U_{N-2}A-U_{N-3}I,
\end{equation}
where $I$ is the unit matrix ,$U_N=(\lambda^N_+-\lambda^N_-)/
(\lambda_+-\lambda_-)$, and $\lambda_{\pm}=(t-z_0/z\pm \sqrt{(t-z_0/z)^2-4})/2$.
Substituting Eq.(10) into Eq.(8), we reduce Eq.(8) to its simplest form
\begin{equation}
U_{N+1}=0.
\end{equation}

$U_{N+1}$ can be rewritten as
\begin{equation}
U_{N+1}=\sin[(N+1)\phi]/\sin\phi
\end{equation}
for $(t-z_0/z)^2\le 4$. Here $\phi=\arccos[(t-z_0/z)/2]$.
For $(t-z_0/z)^2\ge 4$, $\phi$ becomes $i\theta$, and the trigonometric
functions become hyperbolic functions of $\theta$.

Eq.(11) gives
\begin{equation}
t=2\cos[\pi/(N+1)]+z_0/z.
\end{equation}
In the limit $N\rightarrow\infty$, the reduced bulk temperature
$t_B$ is obtained as
\begin{equation}
t_B=2+z_0/z.
\end{equation}

Fig.1 shows the reduced transition temperature vs. layer number $L$.
It can be seen that the film transition tempeature is lower than the bulk
one, i.e., the film disorders at a lower temperature than the bulk ones.
Throughout this paper, we take $J$ as the unit of energy.

In the above discussions, we have assumed the surface exchange constants $J_s$
are the same as the bulk exchange constants $J$. Now we consider a $(l,n,l)$
film consisting $l$ top surface layers, $n$ bulk surface layers and
$l$ bottom surface layers, and assume that the exchange constants in a surface
layer is denoted by $J_s$ and that in a bulk layer or between successive
layers by $J$. In this case the total transfer matrix $R$ becomes
\begin{equation}
R=P^{l-1}Q^nP^{l-1},
\end{equation}
where the matrix $P$ and $Q$ are
\begin{equation}
P=\left(\matrix{t-4J_s/J&-1\cr 1&0}\right), Q=\left(\matrix{t-4&-1\cr 1&0}\right).
\end{equation}
Here we have assumed that the spins lie on a simple cubic lattice, i.e.,
$z_0=4,z=1$. Since $Det(P)=Det(Q)=1$,
the matrix $P^{l-1}$ and $Q^n$ can be linearalized in
analogous to Eq.(10). In this case, the nonlinear Eq.(7)
reduces to
\begin{equation}
R_{11}(t-4J_s/J)^2+(R_{12}-R_{21})(t-4J_s/J)-R_{22}=0.
\end{equation}

The numerical results for transition
temperatures of a magnetic $(l,n,l)$ film as a function of $J_s/J$
are shown in Fig.2. It is can be seen that the transition
temperature increase as $J_s/J$ increases. For $J_s/J\le 1$,
the transition temperatures in film $(10,10,10)$ are nearly
equal to the bulk temperature $T_B$.
It is interesting that the transtion
temperature increases linearly with the
increase of $J_s/J$ when $J_s/J$ is large enough.
We can also see that the transition temperature increases as layer number
increases.

\section{Phase transitions in Ising magnetic superlattices}
The $(l,n)$ superlattice structure we study is formed from two
types of atoms.  In each elementary unit with layer indices $i=1,2...l+n$,
there are $l$ atomic layers of type $A$ and $n$ atomic layers of type $B$.
The interlayer exchange constants are given by $J_a$ and $J_b$, whereas
the exchange constants between different layers is described by $J$.
For the above model of the superlattice, the transfer matrix $M_i$(Eq.(5))
reduce to different types of matrix
\begin{equation}
A=\left(\matrix{X_A&-1\cr 1&0}\right), B=\left(\matrix{X_B&-1\cr 1&0}\right),
\end{equation}
where $X_A=t-4J_A/J$ and $X_B=t-4J_B/J$.
From Eq.(4), we can obtain the following equation as
\begin{equation}
{ m_{l+n+2}\choose m_{l+n+1}} = R{m_2\choose m_1},
\end{equation}
where
\begin{equation}
R=AB^nA^{l-1}
\end{equation}
is the total transfer matrix.

Due to the periodicity of the superlattice, we know $m_{l+n+2}=m_2$ and
$m_{l+n+1}=m_1$. Then from Eq.(19), we get
\begin{equation}
Det(R)-Tr(R)+1=0.
\end{equation}
It can be easily seen that $Det(A)=Det(B)=Det(R)=1$. Then the Eq.(21)
reduces to the simplest form
\begin{equation}
Tr(R)=2.
\end{equation}

Actually, the above equation is a general equation for phase transition
temperature of superlattices. It is valid
for arbitrary exchange constants $J_{ij}$. The nonlinear eqution for films
are dependent on both diagonal and nondiagonal terms of the total transfer
matrix $R$, while the nonlinear equation for superlattices are only depend on
diagonal terms of $R$.

For the total transfer matrix $R=AB^nA^{l-1}$, we get
\begin{equation}
Tr(A^lB^n)=2.
\end{equation}
The matrix $A^l$ and $B^n$ can be linearlized as
\begin{eqnarray}
A^l&=&E_lA-E_{l-1}I\nonumber\\
B^n&=&F_nB-F_{n-1}I,
\end{eqnarray}
where $E_l=(\alpha_+^{l}-\alpha_{-}^{l})/(\alpha_+-\alpha_-)$, $F_n=(\beta_+^n-\beta_-^n)
/(\beta_+-\beta_-)$, $\alpha_\pm=(X_A\pm\sqrt{X^2_A-4})/2$ and
$\beta_\pm=(X_B\pm\sqrt{X^2_B-4})/2$. From Eq.(23), we get the equation
\begin{equation}
(E_lX_a-E_{l-1})(F_nX_b-F_{n-1})-2E_lF_n+E_{l-1}F_{n-1}=2.
\end{equation}
For $l=1$, $n=1$, the above equation reduces to
\begin{equation}
X_aX_b=4,
\end{equation}
which is identical with the result of Ref.(2) and (29).
Next we numerically calculate
the phase transition temperatures from Eq.(25).

In figure 3, we have shown the results for the superlattices
$(3,1),(2,2),(10,5)$ and $(20,20)$. The transition temperature is plotted as a function of $J_A/J$.
For $J_A/J<1$, the transition temperature is smaller that the bulk transition temperature. For $J_A/J=1$, the transition temperature of the superlattice is independent of $m$ and $n$.
and are equal to the bulk temperature $T_B$ as expected.
On the other hand, for $J_A/J>1$, the transition temperature
is greater than the bulk temperature $T_B$. The transition temperature increases with the layer number in one unit cell and approaches $T_B$ asymptotically as the number become large.
The transition temperature increases nearly linearly
with $J_A/J$ when $J_A/J$ is large enough. The layer number of superlattices
$(3,1)$ and $(2,2)$ are same, but the transition temperatures are different.
For $J_A/J<1$, the transition temperatures of superlattice $(2,2)$ are
larger than those of superlattice $(3,1)$. In contrary to this,
the transition temperatures of superlattice $(2,2)$ are
smaller than those of superlattice $(3,1)$ for $J_A/J>1$.  

\section{Discussions}
In summary, we have studied phase transitions in Ising
magnetic films and superlattices with the framework of mean
field theory. By transfer matrix method, we have derived
two general nonlinear equations for phase transition temperatures
in Ising films and superlattices, respectively. The transition temperatures
as a function of exchange interaction constants are calculated.
In addition, the equations can be easily solved and
the parameters involved can be adjusted at will.

\vspace{2cm}
{\bf Figure Captions}\\
Fig.1, Transition temperatures of a uniform film
as a function of layer number $L$.\\
Fig.2, Transition temperatures of a
magnetic film $(l,n,l)$ as a function of $J_s/J$.\\
Fig.3, Transition temperatures of a magnetic
superlattice $(l,n)$ as a function of $J_A/J$.

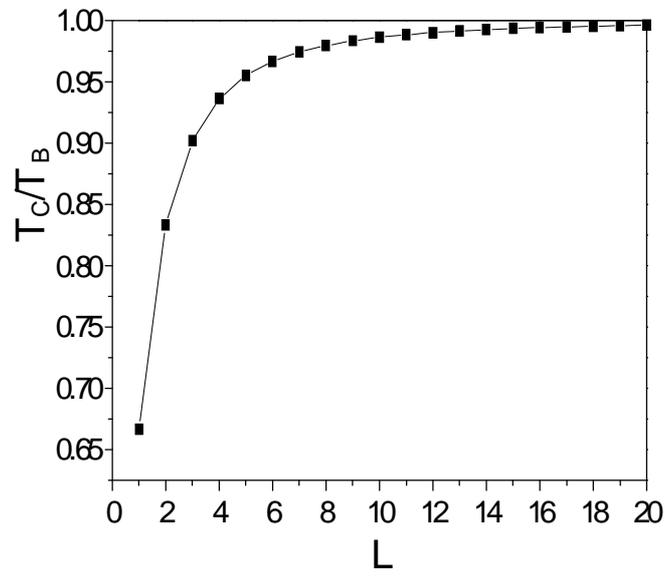
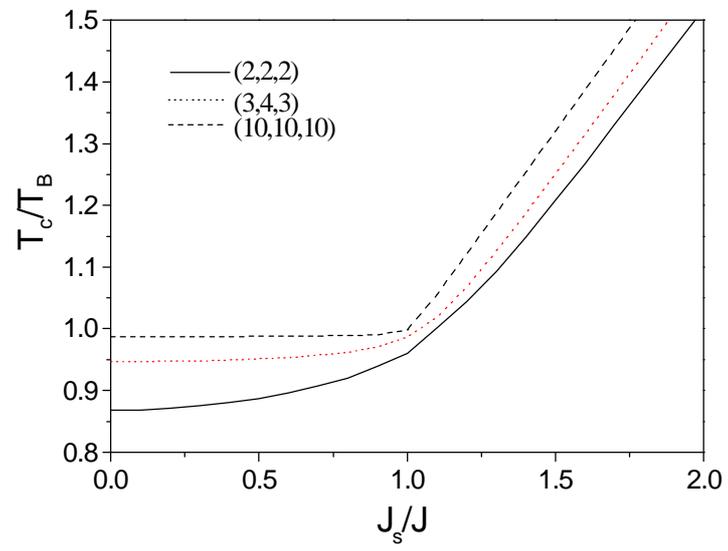
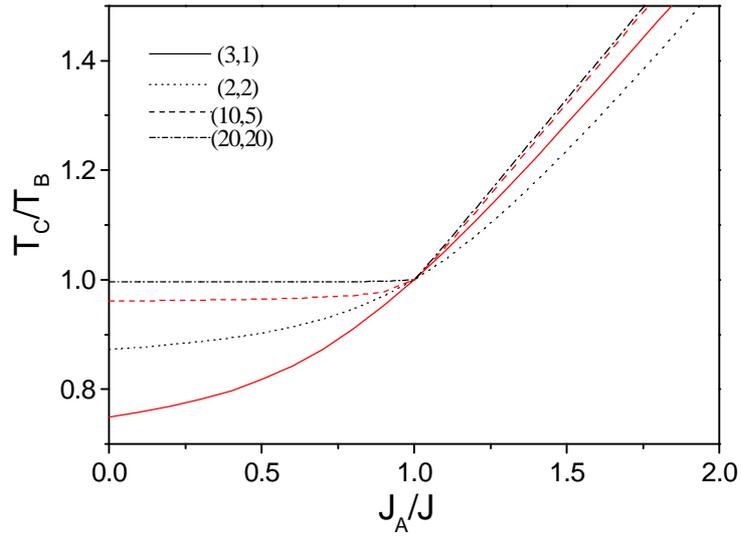
\end{document}